%% file: IONconf_main.tex
\documentclass[letterpaper,times]{IONconf}
\usepackage{url}

\usepackage{natbib}

\usepackage[hidelinks]{hyperref}

\usepackage[printonlyused]{acronym}
\newacro{5g} [5G] {fifth generation}
\newacro{6g} [6G] {sixth generation}
\newacro{mimo} [MIMO] {multiple-input-multiple-output}
\newacro{gnss} [GNSS] {global navigation satellite system}
\newacro{bs} [BS] {base station}
\newacro{los} [LOS] {line-of-sight}
\newacro{ls} [LS] {least-squares}
\newacro{ofdm} [OFDM] {orthogonal frequency-division multiplexing}
\newacro{awgn} [AWGN] {additive white Gaussian noise}
\newacro{crlb} [CRLB] {Cram\'er-Rao lower bound}
\newacro{fim} [FIM] {Fisher information matrix}
\newacro{rmse} [RMSE] {root mean square error}
\newacro{snr} [SNR] {signal-to-noise ratio}
\newacro{bcs} [BCS] {body coordinate system}
\newacro{lcs} [LCS] {local reference coordinate system}
\newacro{gcs} [GCS] {global coordinate system}
\newacro{aoa} [AoA] {angle-of-arrival}
\newacro{dd} [DD] {double differencing}
\newacro{tdma} [TDMA] {time-division multiple access}
\newacro{mclambda} [MC-LAMBDA] {multivariate constrained LAMBDA}
\usepackage{color}

\usepackage{tikz} 
\usepackage[utf8]{inputenc}
\usepackage{pgfplots} 
\usepackage{tabu,longtable}
\usepackage{diagbox}
\usepackage{threeparttable}
\usepackage{makecell}
\usepackage{graphicx}
\usepackage{caption}
\usepackage{amsthm} 

\usepackage{units} 
\usepackage{bm} 
\usepackage{amsmath} 
\usepackage{amssymb} 
\include{macros} 
\usepackage{setspace} 
\title{Attitude Determination in Urban Canyons: A Synergy between GNSS and 5G Observations}

\author{
Pinjun~Zheng,
Xing~Liu,
Tarig~Ballal,
and~Tareq~Y.~Al-Naffouri
\vspace{1mm} \\%
\textit{King Abdullah University of Science and Technology (KAUST)}%
    }

\begin{document}

\maketitle

\section*{biography}


\biography{Pinjun~Zheng}{is a PhD student with the Electrical and Computer Engineering Program, King Abdullah University of Science and Technology (KAUST), Thuwal, Saudi Arabia. His current research focuses on mmWave/THz localization and communications.}

\biography{Xing~Liu}{is a postdoctoral researcher with the Electrical and Computer Engineering Program, King Abdullah University of Science and Technology (KAUST), Thuwal, Saudi Arabia. His current research focuses on precise GNSS positioning and attitude determination.}

\biography{Tarig~Ballal}{is a research scientist with the Electrical and Computer Engineering Program, King Abdullah University of Science and Technology (KAUST), Thuwal, Saudi Arabia. His research interests lie in the areas of signal and image processing, localization and tracking, in addition to GNSS positioning and attitude determination.}

\biography{Tareq~Y.~Al-Naffouri}{is a professor with the Electrical and Computer Engineering Program, King Abdullah University of Science and Technology (KAUST), Thuwal, Saudi Arabia. His current research focuses on the areas of sparse, adaptive, and statistical signal processing and their applications, as well as machine learning and network information theory.}

\section*{Abstract}

This paper considers the attitude determination problem based on the \ac{gnss} and \ac{5g} measurement fusion to address the shortcomings of standalone \ac{gnss} and \ac{5g} techniques in deep urban regions.
The tight fusion of the \ac{gnss} and the \ac{5g} observations results in a unique hybrid integer- and orthonormality-constrained optimization problem.
To solve this problem, we propose an estimation method consisting of the steps of float solution computation, ambiguity resolution, and fixed solution computation.
Numerical results reveal that the proposed method can effectively improve the attitude determination accuracy and reliability compared to either the pure \ac{gnss} solution or the pure \ac{5g} solution.

\section{INTRODUCTION}
With numerous emerging technologies flourishing in areas such as wireless communications, robotics, and artificial intelligence, many facets of human society are benefiting from intelligent location/attitude-aware services \citep{Di2014Location,Alletto2016Indoor,Jiang2021OmniTrack}.
Beyond the location information, determining the user's attitude information becomes more and more important nowadays~\citep{Douik2020Precise,Liu2022Instantaneous}.
GNSS attitude determination has been widely studied in literature \citep{giorgi2010carrier, teunissen2012affine, 9837939, 9110131, liu2022gnss,liu2023gnss}. 
Compared to other existing attitude determination techniques, such as inertial sensors, \ac{gnss} attitude determination enjoys the advantages of being driftless, power efficient, low cost, and requiring minor maintenance.
However, in deep urban regions, \ac{gnss} performance usually degrades, falling short of meeting the accuracy requirements of many applications. In such environments, the surrounding buildings can block, weaken, reflect, and diffract the \ac{gnss} signals, which may result in an insufficient number of visible satellites and/or severe observation errors due to multipath effects~\citep{groves2013gnss, liu2019gnss}.
Integrating GNSS and inertial navigation systems is one way to mitigate these limitations in \ac{gnss}-deprived environments~\citep{angrisano2010gnss, falco2017loose, wen2021factor}.

Recently, as the exploration of the \ac{5g} and \ac{6g} wireless communication systems continue to move forward, the provision of high-precision localization services (comprised of the user's position and attitude) is regarded as an increasingly crucial feature of \ac{5g}/\ac{6g} systems~\citep{tr2370086}.
Over the years, both the implemented algorithms and the theoretical bounds for achieving localization within \ac{5g}/\ac{6g} systems have been extensively studied~\citep{del2018Survey,Abu2018Error}.
Several works on attitude estimation in mmWave/THz \ac{mimo} systems have been reported in the literature, indicating that an attitude estimation accuracy of \unit[0.1]{$^\circ$}--\unit[1]{$^\circ$} can be achieved in a typical mmWave/THz \ac{mimo} system~\citep{Zheng2022Coverage}.
However, most of these works rely on the availability of signals from enough \ac{5g}/\ac{6g} \acp{bs}, which may not be feasible in dense urban areas.
Moreover, calibration errors due to the geometric information of the \ac{5g}/\ac{6g} \acp{bs} can result in drastic performance loss~\citep{Zheng2023Misspecified}.
To address these limitations, an appealing idea is to integrate measurements from 5G/6G systems together with \ac{gnss} observations.
Some efforts have been put into developing hybrid \ac{5g}/\ac{6g}-\ac{gnss} localization systems, which is a practical solution thanks to the ubiquitous wireless radio resources in urban areas.
More specifically, \ac{5g} observations have been demonstrated to be useful in \ac{gnss}-deprived environments because they can not only improve positioning availability in extreme scenarios but also enhance the estimation accuracy in less severe environments~\citep{Zheng20235G}.
Nonetheless, as an important component of the user state, attitude estimation is rarely discussed in hybrid \ac{5g}/\ac{6g}-\ac{gnss} localization systems.

In this work, we develop a novel attitude determination method based on hybrid \ac{5g} and \ac{gnss} observations, including GNSS pseudo-ranges, GNSS carrier phases, and \ac{5g} \acp{aoa}.
By rigorously incorporating these observations, we formulate a constrained weighted least-squares problem to estimate the integer ambiguities of the \ac{gnss} carrier phase observations and the rotation matrix of the user platform.
Initially, we neglect all the constraints on the underlying unknowns and obtain a closed-form float solution.
Then, the integer ambiguities in the \ac{gnss} carrier phase observations are resolved using the popular \ac{mclambda} method. 
Finally, the rotation matrix of the user platform is updated after the integer ambiguity resolution is completed to obtain the fixed solution.
We will show that the hybrid method can considerably improve performance compared to the standalone \ac{gnss} and \ac{5g} solutions. This improvement is attributed to the tight incorporation of the hybrid observations and the well-designed weight matrices which characterize the dispersion of the observations and the initial closed-form solution.

\section{HYBRID GNSS-5G ATTITUDE DETERMINATION SYSTEM}
\begin{figure}[t]
    \centering
    \includegraphics[width=6in]{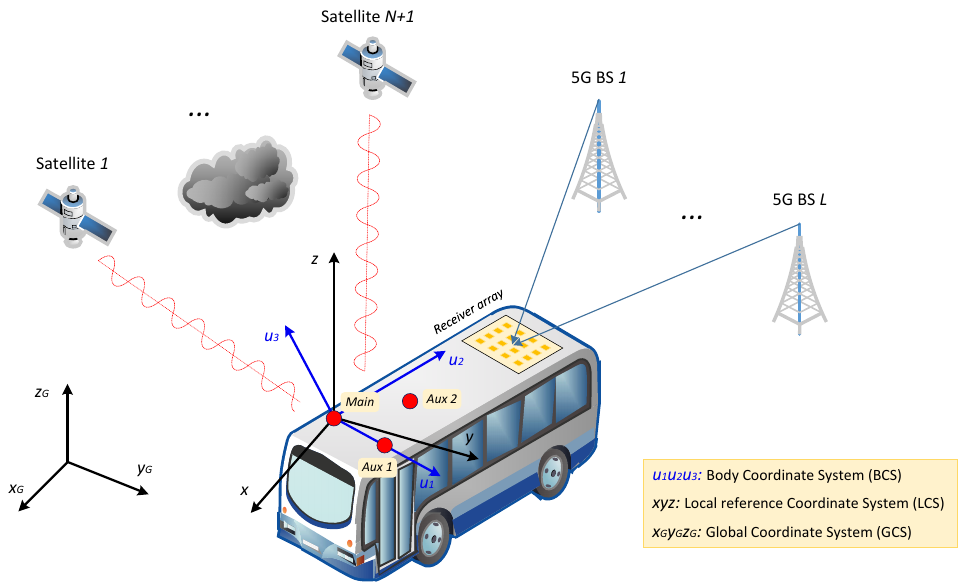}
    \caption{ 
        Illustration of a \ac{5g}-aided \ac{gnss} attitude determination system with $N+1$ \ac{gnss} satellites and $L$ \ac{5g} \acp{bs}. 
        $Main$, $Aux 1$, and $Aux 2$ are the main and the auxiliary \ac{gnss} antennas that define the two baselines, and the BCS is built accordingly.
        A multi-antenna 5G receiver is equipped to resolve the \ac{aoa} from different 5G \acp{bs} to the receiver array.
      }
    \label{fig_1}
  \end{figure}

\subsection{System Description}
To take advantage of both \ac{gnss} and 5G measurements, a user platform integrating $M+1$ \ac{gnss} antennas (which form $M$ baselines) and a multi-antenna 5G receiver is adopted, as shown in Fig.~\ref{fig_1}.
The \ac{gnss} observables are the pseudo-range and carrier phase measurements obtained by tracking $N+1$ satellites.
These observations are contaminated with errors such as clock biases, instrumental delays, atmospheric delays, and multipath~\citep{teunissen2017springer}.
By \ac{dd} the pseudo-range and carrier phase observations (i.e. constructing the differences between observations collected at two \ac{gnss} antennas from two different satellites), the clock biases, instrumental delays, and atmospheric delays become negligible~\citep{teunissen2012gps}.
The 5G receiver, on the other hand, can receive 5G pilot signals from $L$ \acp{bs} (with known positions) and resolve the \acp{aoa} of different \acp{bs} by applying a channel estimation process~\citep{Shahmansoori2018Position}. 
Hence, we take the estimated \acp{aoa} as the 5G observations, whose estimation errors are related to the distance between the specific \ac{bs} and the receiver, the signal power, the noise level, etc~\citep{Zheng2023JrCUP}.
The pilot signals transmitted from different \acp{bs} are required to be orthogonal in time and/or frequency to avoid interference between channels.

As depicted in Fig.~\ref{fig_1}, there are 3 different coordinate systems, namely, \ac{bcs} $u_1u_2u_3$, \ac{lcs} $xyz$, and \ac{gcs} $x_Gy_Gz_G$.
The receiver's \ac{bcs} is built as follows: 
the first axis $u_1$ is aligned with the first baseline ($Main-Aux1$), the second axis $u_2$ is perpendicular to the first, lying in the plane formed by the first two baselines, and the third body axis $u_3$ is directed so that $u_1u_2u_3$ forms a right-handed orthogonal frame.
In addition, the \ac{lcs} $xyz$ is defined as keeping the same attitude but a different origin as the \ac{gcs} (the origins of the \ac{lcs} and the \ac{bcs} coincide). The attitude of the user is represented by a rotation matrix $\Rm\in\mathrm{SO(3)}$, where $\mathrm{SO(3)}$ denotes the special orthogonal group of 3D rotation matrices as
$
  \mathrm{SO}(3) = \{\Rm|\Rm^\TT\Rm=\mathbf{I}_3,\det(\Rm) = 1\}.
$
The rotation matrix transforms a vector/matrix from the \ac{bcs} to the \ac{lcs}.
Suppose $\tv^{\Bt}$ and $\tv^{\Lt}$ represent the coordinates of the same vector expressed in the \ac{bcs} and the \ac{lcs} respectively, then they are related through the rotation matrix $\Rm$ as 
\begin{equation}\label{eq_R}
    \tv^\Lt = \Rm\tv^\Bt.
\end{equation}

\subsection{Observation Model}
\subsubsection{GNSS Observations}
Considering a set of $2N$ \ac{dd} pseudo-range and carrier phase observations obtained at two \ac{gnss} antennas (i.e., a single baseline) via tracking the signals from $N+1$ satellites, 
the functional model is given by 
\begin{equation}
    \yv = \Am\zv + \Gm\bv + \bm{\epsilon},\quad \Qm(\yv) = \Sigmam_{\bm{\epsilon}}, 
\end{equation}
where $\yv\in\mathbb{R}^{2N}$ is the vector of \ac{dd} pseudo-range and carrier phase observations, $\zv\in\mathbb{Z}^{N}$ is the vector of $N$ integer ambiguities, $\bv\in\mathbb{R}^{3}$ denotes the three baseline coordinates (expressed in \ac{gcs}), and $\bm{\epsilon}\in\mathbb{R}^{2N}$ describes the unmodeled errors.
Here, $\Am\in\mathbb{R}^{2N\times N}$ is the design matrix that contains the carrier wavelengths, while $\Gm\in\mathbb{R}^{2N\times 3}$ is the matrix formed by the combination of unit line-of-sight vectors.
Furthermore, we assume $\bm{\epsilon}$ is an \ac{awgn} with zero mean and covariance matrix $\Sigmam_{\bm{\epsilon}}$.
Hence, the dispersion on $\yv$, denoted as $\Qm(\yv)$, is characterized by the covariance matrix $\Sigmam_{\bm{\epsilon}}$.

Since we consider a configuration with multiple \ac{gnss} antennas, the observations from different baselines are collected to determine the attitude of the user.\footnote{At least three nonlinear antennas (i.e., two baselines) are necessary to estimate the full attitude of a platform~\citep{teunissen2012gps}.}
For $M+1$ \ac{gnss} antennas tracking the same $N+1$ satellites simultaneously, the concatenated observation equation can be formulated as 
\begin{equation}\label{eq_Y}
    \Ym = \Am\Zm + \Gm\Bm + \Xim_1,\quad \Qm(\mathrm{vec}(\Ym)) = \Pm_M\otimes\Sigmam_{\bm{\epsilon}},
\end{equation}
where each column of $\Ym\in\mathbb{R}^{2N\times M}$, $\Zm\in\mathbb{Z}^{N\times M}$, and $\Bm\in\mathbb{R}^{3\times M}$ contains the pseudo-range and carrier phase observations, the integer ambiguities, and the baseline coordinates for each of the baselines, respectively.
The notation $\mathrm{vec}(\cdot)$ stands for the operation of vectorizing a matrix into a vector.
Assuming the observations from different baselines keep the same covariance matrix $\Sigmam_{\bm{\epsilon}}$, the covariance matrix of whole observables can be formulated as $\Qm(\mathrm{vec}(\Ym))=\Pm_M\otimes\Sigmam_{\bm{\epsilon}}$, where $\otimes$ indicates the Kronecker product and $\Pm_M\in\mathbb{R}^{M\times M}$ is given by~\citep{teunissen2012gps}
\begin{equation}
    \Pm_M = \begin{bmatrix}
        1 & 0.5 & \cdots & 0.5 \\
        0.5 & 1 & \cdots & 0.5 \\
        \vdots & \vdots & \ddots & \vdots \\
        0.5 & 0.5 & \cdots & 1
    \end{bmatrix}.
\end{equation}
Based on~\eqref{eq_Y}, the attitude of the user can be connected to observables $\Ym$ through the rotation transformation $\Bm = \Rm\Fm$, 
where $\Fm\in\mathbb{R}^{3\times M}$ indicates the baseline coordinates expressed in the \ac{bcs}. $\Fm$ is assumed to be known since we can precisely measure it in advance.
Therefore, we can rewrite~\eqref{eq_Y} as 
\begin{equation}\label{eq_Y1}
    \Ym = \Am\Zm + \Gm\Rm\Fm + \Xim_1,\quad \Qm(\mathrm{vec}(\Ym)) = \Pm_M\otimes\Sigmam_{\bm{\epsilon}}.
\end{equation}
For notational convenience, we use $\Qm_\yv$ and $\Qm_\Ym$ to denote $\Qm(\yv)$ and $\Qm(\mathrm{vec}(\Ym))$, respectively.

\subsubsection{5G Observations}
Consider a down-link scenario for the \ac{5g} transmissions. We assume all $L$ 5G \acp{bs} work in \ac{tdma} mode, where $T$ transmissions of pilot signals are managed by each 5G \ac{bs}.
On the user  side, the 5G receiver is an $J_1\times J_2$ planar antenna array.
Thus the received 5G signals at the user platform for the $t$-th transmission from the $\ell$-th \ac{bs} can be expressed as~\citep{Chen2022Tutorial}
\begin{equation}
    \yv_{\ell,t} = \hv_\ell x_{\ell,t} + \nv\in\mathbb{C}^{J_1J_2\times 1},
\end{equation}
where $\hv_\ell\in\mathbb{C}^{J_1J_2\times 1}$ is the channel vector, $x_{\ell,t}$ is the transmitted symbol with a power constraint $|x_{\ell,t}|^2 = P_\mathrm{T}$, and $\nv\sim\mathcal{CN}(\mathbf{0},\sigma^2\mathbf{I}_{J_1J_2})$ denotes the thermal noise at the receiver.
Utilizing the far-field model, the $i$-th entry of the channel vector $\hv_\ell$ is given by ~\citep{Chen2022Tutorial}
\begin{equation}
    h_{\ell,i} = \alpha_\ell e^{-j\frac{2\pi f}{c}\pv_{\Rt,i}^\TT\tv_\ell},
\end{equation}
where $\alpha_\ell$ denotes the complex gain of the channel between the $\ell$-th \ac{bs} and the receiver, $\pv_{\Rt,i}\in\mathbb{R}^3$ denotes the position of the $i$-th element of the antenna array at the receiver, and $\tv_\ell\in\mathbb{R}^3$ represents the direction unit vector that coincides with the \ac{aoa} of the signal from the $\ell$-th \ac{bs} to the 5G receiver on the user platform.
Note that both $\pv_{\Rt,i}$ and $\tv_\ell$ are expressed in the \ac{bcs}. 
Besides, $f$ represents the \ac{5g} carrier frequency, $c$ is the speed of light, and $(\cdot)^\TT$ denotes the transpose operation.

Based on the received symbols $\yv_{\ell,t}$, the \ac{aoa} vector $\tv_\ell$ in the \ac{bcs} can be estimated via various channel estimators, such as atomic norm minimization~\citep{He2021Channel} and ESPRIT~\citep{Zheng2023JrCUP}. 
Therefore, we define the 5G observables as 
\begin{equation}\label{eq_Dm}
    \Dm = [\tv_1,\dots,\tv_L]\in\mathbb{R}^{3\times L}.
\end{equation}
The dispersion on $\mathrm{vec}(\Dm)$ can be approximated by the inverse of its \ac{fim}~\citep{kay1993fundamentals,Shahmansoori2018Position}, which we denote as $\Qm_\Dm$.

In this work, we assume the position of the user platform $\pv_\Ut\in\mathbb{R}^3$,  which coincides with the origin of the \ac{lcs} (and the \ac{bcs}), to be known in the \ac{gcs}.\footnote{For example, the user position $\pv_\Ut$ can be estimated using the real-time kinematic technique~\citep{Zheng20235G} or \ac{5g}/6G localization methods~\citep{Shahmansoori2018Position,chen2023multi}, which is not discussed in this paper as it is beyond the scope of this work.}
We further assume the positions of the 5G \acp{bs} in the \ac{gcs} denoted as $\pv_{\Bt,\ell},\ \ell=1,\dots,L$ to be precisely known. 
Since we know the positions of the 5G \acp{bs} and the user platform, we can compute the columns of $\Dm$ expressed in the \ac{gcs} as
\begin{equation}
    \tv_\ell^\Gt = \frac{\pv_{\Bt,\ell} - {\pv}_\Ut}{\|\pv_{\Bt,\ell} - {\pv}_\Ut\|},\quad \ell=1,\dots,L.
\end{equation}
Again, the \ac{lcs} and the \ac{gcs} have the same attitude but only the origins are different, so $\tv_\ell^\Lt=\tv_\ell^\Gt$.
We further define $\Em = [\tv_1^\Lt,\dots,\tv_L^\Lt]=[\tv_1^\Gt,\dots,\tv_L^\Gt]\in\mathbb{R}^{3\times L}$ as the LCS version of $\Dm$. Then, according to~\eqref{eq_R}, the 5G observations are related to the user attitude as 
\begin{equation}\label{eq_D1}
   \Dm = \Rm^\TT \Em + \Xim_2, \quad \Qm(\mathrm{vec}(\Dm)) = \Qm_\Dm,
\end{equation}
where $\Xim_2\in\mathbb{R}^{3\times L}$ combines additive errors.

\subsubsection{Hybrid GNSS-5G Observations}

Combining~\eqref{eq_Y1} and~\eqref{eq_D1}, we have the hybrid observation model for attitude determination as
\begin{align}\label{eq_hybrid}
    &\begin{bmatrix}
        \mathrm{vec}(\Ym) \\
        \mathrm{vec}(\Dm^\TT)
    \end{bmatrix} = 
    \begin{bmatrix}
        \mathbf{I}\otimes\Am & \Fm^\TT\otimes\Gm\\
        \mathbf{0} & \mathbf{I}\otimes\Em^\TT
    \end{bmatrix}
    \begin{bmatrix}
        \mathrm{vec}(\Zm)\\
        \mathrm{vec}(\Rm)
    \end{bmatrix},\\
    &\ \Zm\in\mathbb{Z}^{N\times M},\quad \Rm\in\mathrm{SO(3)}. \notag
\end{align}
Since the \ac{gnss} receiver and the 5G receiver are independent, the covariance matrix of the hybrid observables is given by
\begin{equation}\label{eq_hybridQ}
   \Qm_{\Ym,\Dm} = \Qm([\mathrm{vec}(\Ym)^\TT,\mathrm{vec}(\Dm^\TT)^\TT]^\TT) = \mathrm{blkdiag}(\Qm_\Ym,\Qm_\Dm),
\end{equation}
where $\mathrm{blkdiag}(\cdot,\cdot)$ represents constructing a block diagonal matrix from input sub-matrices.

\section{ATTITUDE DETERMINATION METHODOLOGY}
Based on~\eqref{eq_hybrid} and~\eqref{eq_hybridQ}, the objective function of the hybrid attitude determination problem can be formulated as
\begin{align}\label{eq_WLS}
    \min_{\Zm\in\mathbb{Z}^{N\times M},\ \Rm\in\mathrm{SO(3)}} \left\|
    \begin{bmatrix}
        \mathrm{vec}(\Ym) \\
        \mathrm{vec}(\Dm^\TT)
    \end{bmatrix} - 
    \begin{bmatrix}
        \mathbf{I}\otimes\Am & \Fm^\TT\otimes\Gm\\
        \mathbf{0} & \mathbf{I}\otimes\Em^\TT
    \end{bmatrix}
    \begin{bmatrix}
        \mathrm{vec}(\Zm)\\
        \mathrm{vec}(\Rm)
    \end{bmatrix}
    \right\|_{\Qm_{\Ym,\Dm}^{-1}}^2,
\end{align}
where $\|\cdot\|_\Qm=(\cdot)^\TT\Qm^{-1}(\cdot)$. 
The optimization in~\eqref{eq_WLS} is a constrained weighted \ac{ls} problem. Due to the presence of the integer constraints, there exists no closed-form solution for \eqref{eq_WLS}. We can first disregard the integer and orthonormality constraints to calculate a closed-form solution for~\eqref{eq_WLS}, i.e.,a \textit{float solution}, and then pursue the final constrained solution using a search procedure around the float solution. In the search process, the ambiguities can be resolved based on the integer least-squares (ILS) principle~\citep{teunissen1994integer}.
Then, the estimated ambiguities are used to compute the \textit{fixed solution}.

\subsection{The Float Estimator}
\label{sec:flaot}
Without the integer and the orthonormality constraints, based on the \ac{ls} principle, the normal equations of~\eqref{eq_WLS} are given by
\begin{align}
    &\Nm\begin{bmatrix}
        \mathrm{vec}(\hat{\Zm})\\
        \mathrm{vec}(\hat{\Rm})
    \end{bmatrix} = \begin{bmatrix}
        \mathbf{I}\otimes\Am^\TT & \mathbf{0}\\
        \Fm\otimes\Gm^\TT & \mathbf{I}\otimes\Em
    \end{bmatrix}\Qm_{\Ym,\Dm}^{-1}
    \begin{bmatrix}
        \mathrm{vec}(\Ym) \\
        \mathrm{vec}(\Dm^\TT)
    \end{bmatrix},\\
    &\Nm = \begin{bmatrix}
        \mathbf{I}\otimes\Am^\TT & \mathbf{0}\\
        \Fm\otimes\Gm^\TT & \mathbf{I}\otimes\Em
    \end{bmatrix}\Qm_{\Ym,\Dm}^{-1}
    \begin{bmatrix}
        \mathbf{I}\otimes\Am & \Fm^\TT\otimes\Gm\\
        \mathbf{0} & \mathbf{I}\otimes\Em^\TT
    \end{bmatrix}.
\end{align} 
That results in the following float solution:
\begin{equation}\label{eq_floatsolu}
	\begin{bmatrix}
        \mathrm{vec}(\hat{\Zm})\\
        \mathrm{vec}(\hat{\Rm})
    \end{bmatrix} = \Nm^{-1}\begin{bmatrix}
        \mathbf{I}\otimes\Am^\TT & \mathbf{0}\\
        \Fm\otimes\Gm^\TT & \mathbf{I}\otimes\Em
    \end{bmatrix}\Qm_{\Ym,\Dm}^{-1}
    \begin{bmatrix}
        \mathrm{vec}(\Ym) \\
        \mathrm{vec}(\Dm^\TT)
    \end{bmatrix}.
\end{equation}

The covariance matrix of this float solution is obtained by inversion of the normal matrix, i.e.,
\begin{equation}
	\Qm_{\mathrm{hybrid}} = 
    \begin{bmatrix}
        \Qm_{\hat{\Zm}} & \Qm_{\hat{\Zm}\hat{\Rm}} \\
        \Qm_{\hat{\Rm}\hat{\Zm}} & \Qm_{\hat{\Rm}}
    \end{bmatrix} = \Nm^{-1}.
\end{equation}

Now, we can show the superiority of the hybrid solution in~\eqref{eq_floatsolu} by analyzing the covariance matrix $\Qm_{\mathrm{hybrid}}$. Defining $\Mm_1 = [\mathbf{I}\otimes\Am,\ \Fm^\TT\otimes\Gm]$ and $\Mm_2 = [\mathbf{0},\ \mathbf{I}\otimes\Em^\TT]$, we have
\begin{equation}
	\Qm_{\mathrm{hybrid}}^{-1} = 
	\begin{bmatrix}
		\Mm_1^\TT & \Mm_2^\TT
	\end{bmatrix}
	\begin{bmatrix}
		\Qm_\Ym^{-1} & \mathbf{0} \\
		\mathbf{0} & \Qm_\Dm^{-1}
	\end{bmatrix}
	\begin{bmatrix}
		\Mm_1 \\
		\Mm_2
	\end{bmatrix} = 
	\underbrace{\Mm_1^\TT\Qm_\Ym^{-1}\Mm_1}_{\Qm_\mathrm{GNSS}^{-1}} + \underbrace{\Mm_2^\TT\Qm_\Dm^{-1}\Mm_2}_{\Qm_\mathrm{5G}^{-1}},
\end{equation}
where $\Qm_\mathrm{GNSS}$ and $\Qm_\mathrm{5G}$ represent the covariance matrices of the standalone \ac{gnss} and \ac{5g} float solutions, respectively.
Since the covariance matrices $\Qm_{\mathrm{hybrid}}$, $\Qm_\mathrm{GNSS}$, and $\Qm_\mathrm{5G}$ are positive semidefinite, we have
\begin{align}\label{eq_compareCov}
	\Qm_{\mathrm{hybrid}}^{-1} \succeq \Qm_\mathrm{GNSS}^{-1},\qquad\Qm_{\mathrm{hybrid}}^{-1} \succeq \Qm_\mathrm{5G}^{-1},
\end{align}
where $\Am\succeq\Bm$ stands for matrix $\Am-\Bm$ is positive semidefinite.
The relationships in~\eqref{eq_compareCov} immediately yield~\citep{horn2012matrix}
\begin{equation}
	\Qm_{\mathrm{hybrid}} \preceq \Qm_\mathrm{GNSS},\qquad \Qm_{\mathrm{hybrid}} \preceq \Qm_\mathrm{5G}.
\end{equation}
Hence we have
\begin{equation}\label{eq_TrQ}
	\mathrm{Tr}(\Qm_{\mathrm{hybrid}}) \leq \mathrm{Tr}(\Qm_\mathrm{GNSS}),\qquad \mathrm{Tr}(\Qm_{\mathrm{hybrid}}) \leq \mathrm{Tr}(\Qm_\mathrm{5G}),
\end{equation}
where $\mathrm{Tr}(\cdot)$ returns the trace of a square matrix. Here,~\eqref{eq_TrQ} stipulates that the error covariance of the hybrid solution~\eqref{eq_floatsolu} is lower than that of either of the standalone \ac{gnss}/\ac{5g} solutions.

\subsection{The Multivariate Constrained Estimator}
Following a similar procedure to the standard method for GNSS attitude determination~\citep{giorgi2010carrier}, the objective function can be decomposed based on the float solution in~\eqref{eq_floatsolu} as follows:
\begin{equation} \label{eq:decomp}
\begin{aligned}
    \min_{\Zm\in\mathbb{Z}^{N\times M},\ \Rm\in\mathrm{SO(3)}}\quad & \left\|
    \begin{bmatrix}
        \mathrm{vec}(\Ym) \\
        \mathrm{vec}(\Dm^\TT)
    \end{bmatrix} - 
    \begin{bmatrix}
        \mathbf{I}\otimes\Am & \Fm^\TT\otimes\Gm\\
        \mathbf{0} & \mathbf{I}\otimes\Em^\TT
    \end{bmatrix}
    \begin{bmatrix}
        \mathrm{vec}(\Zm)\\
        \mathrm{vec}(\Rm)
    \end{bmatrix}
    \right\|_{\Qm_{\Ym,\Dm}^{-1}}^2\\
    = &\left\|
    \begin{bmatrix}
        \mathrm{vec}(\Ym) \\
        \mathrm{vec}(\Dm^\TT)
    \end{bmatrix} - 
    \begin{bmatrix}
        \mathbf{I}\otimes\Am & \Fm^\TT\otimes\Gm\\
        \mathbf{0} & \mathbf{I}\otimes\Em^\TT
    \end{bmatrix}
    \begin{bmatrix}
        \mathrm{vec}(\hat \Zm)\\
        \mathrm{vec}(\hat \Rm)
    \end{bmatrix}
    \right\|_{\Qm_{\Ym,\Dm}^{-1}}^2\\
    &\qquad + \mathop {\min }\limits_{\Zm\in\mathbb{Z}^{N\times M}}  \left(\left\| {\mathrm{vec} \! \left( \Zm  - \hat \Zm \right)} \right\|_{\Qm_{\hat{\Zm}}^{-1}}^2+
\mathop {\min }\limits_{\Rm\in\mathrm{SO(3)}}\left\| {\mathrm{vec}\left(\Rm - \hat \Rm \left( \Zm \right) \right)} \right\|_{\Qm_{{\hat{\Rm}}\left( \Zm\right)}^{-1}}^2 \right).
\end{aligned}
\end{equation}
If the integer ambiguities $\Zm$ are known, the float estimate of $\Rm$ can be updated as~\citep{giorgi2010carrier} 
\begin{equation}
    \mathrm{vec}(\hat{\Rm}(\Zm)) = \mathrm{vec}(\hat{\Rm}) - \Qm_{\hat{\Rm}\hat{\Zm}}\Qm_{\hat{\Zm}}^{-1}\mathrm{vec}(\hat{\Zm}-\Zm).
\end{equation}
The covariance matrix of the updated solution is given by  
\begin{equation}
    \Qm_{\hat{\Rm}(\Zm)} = \Qm_{\hat{\Rm}} - \Qm_{\hat{\Rm}\hat{\Zm}} \Qm_{\hat{\Zm}}^{-1} \Qm_{\hat{\Zm}\hat{\Rm}}.
\end{equation}

Note that the first term of the right-hand side of \eqref{eq:decomp} does not depend on $\mathbf{Z}$ and $\mathbf{R}$. 
Consequently, the integer ambiguities can be resolved based on the following minimization problem:
\begin{align} \label{eq:searcj}
   \check \Zm = \mathop {\arg\min }\limits_{\Zm\in\mathbb{Z}^{N\times M}}  \mathcal{C} \!\left ( \Zm \right ),
\end{align}
where
\begin{align}
   \mathcal{C} \!\left ( \Zm \right ) = \left\| {\mathrm{vec} \! \left( \Zm  - \hat \Zm \right)} \right\|_{\Qm_{\hat{\Zm}}^{-1}}^2+
\mathop {\min }\limits_{\Rm\in\mathrm{SO(3)}}\left\| {\mathrm{vec}\left(\Rm - \hat \Rm \left( \Zm \right) \right)} \right\|_{\Qm_{{\hat{\Rm}}\left( \Zm\right)}^{-1}}^2,
\end{align}
while the fixed solution of the rotation matrix is given by
\begin{align}
   \check \Rm  = 
\mathop {\min }\limits_{\Rm\in\mathrm{SO(3)}}\left\| {\mathrm{vec}\left(\Rm - \hat \Rm \left( \check \Zm \right) \right)} \right\|_{\Qm_{{\hat{\Rm}}\left( \Zm\right)}^{-1}}^2.
\end{align}
An efficient integer search strategy, either the search-and-shrink or search-and-expansion algorithm, can be utilized to solve \eqref{eq:searcj} \citep{giorgi2010carrier}.

\section{Performance Evaluation}

\subsection{Evaluation Setup}

\begin{table}[h]
\renewcommand{\arraystretch}{1.5}
    \begin{center}
    \begin{threeparttable}
    \caption{The relative positions of the \ac{5g} \acp{bs} (m)}
    \label{tab1}
    \setlength{\tabcolsep}{1.8mm}{
        \begin{tabular}{|c|c|c|c|c|c|c|c|c|}
        \hline
        \  & BS~1 & BS~2 & BS~3 & BS~4 & BS~5 & BS~6 & BS~7 & BS~8\\
        \hline\hline
        $x$ & 10 & -10 & 5 & -5 & 15 & -15 & 0 & 0   \\
        \hline
        $y$ & 10 & -10 & 10 & -10 & 0 & 0 & 15 & -15 \\
        \hline
        $z$ & 10 & 10 & 15 & 15 & 10 & 10 & 10 & 10 \\
        \hline
        \end{tabular}
    }
    \end{threeparttable}
    \end{center}
\end{table}

The simulations are implemented using the real data of satellite orbit information and the assumed user position and attitude. We generate GNSS observations based on only GPS constellation. Without further specification, the default number of the tracked GPS satellites that we simulate is 5. We use 4 \ac{gnss} antennas which form 3 baselines with unit vectors $\bv_1=[1,0,0]^\TT$, $\bv_2=[0,1,0]^\TT$, and $\bv_3=[0,0,1]^\TT$, in units of meters.
We set the standard deviation of the carrier-phase measurements equal to a value $\sigma$ and that of the pseudo-range data equal to $100\sigma$.
By default, we set $\sigma=\unit[0.001]{m}$.
The position of the $i$-th \ac{5g} \acp{bs} is denoted as $\pv_i=\pv_\Ut + \Delta\pv_i$.
For trials under different number of \ac{5g} \acp{bs}, the \ac{bs}'s relative positions to the user platform, $\Delta\pv_i,\ i=1,...,8,$ are picked from Table~\ref{tab1}. The \ac{5g} \acp{bs} transmit pilots at a carrier frequency of $\unit[28]{GHz}$ with $\unit[300]{MHz}$ bandwidth, and the size of the receiver array is set as $5\times 5$ with half-wavelength spacing. 
By default, we set the number of \ac{5g} transmissions at a single observation as $T=128$, and the average transmission power is $\unit[17]{dBm}$.
We assume the received \ac{5g} signals are contaminated by an \ac{awgn} with a noise power spectral density of $\unit[-174]{dBm/Hz}$.
Therefore, the covariance matrix of the \ac{5g} observations $\Dm$ in~\eqref{eq_Dm} can be evaluated by the inverse of the \ac{fim}, as specified in~\citep{Zheng2022Coverage,Zheng2023JrCUP}.

Throughout the simulation studies, the proposed method is compared with the standalone \ac{gnss} and the standalone \ac{5g} methods, which use the pure \ac{gnss} and the pure \ac{5g} observations, respectively.
For standalone \ac{gnss} estimation, we use the method proposed in~\citep{giorgi2010carrier}.
For standalone \ac{5g} estimation, we use the method proposed in~\cite[Sec. V-C]{Zheng2022Coverage}.

\subsection{Simulation Results}

\begin{figure}[t]
    \centering
    \includegraphics[width=6.5in]{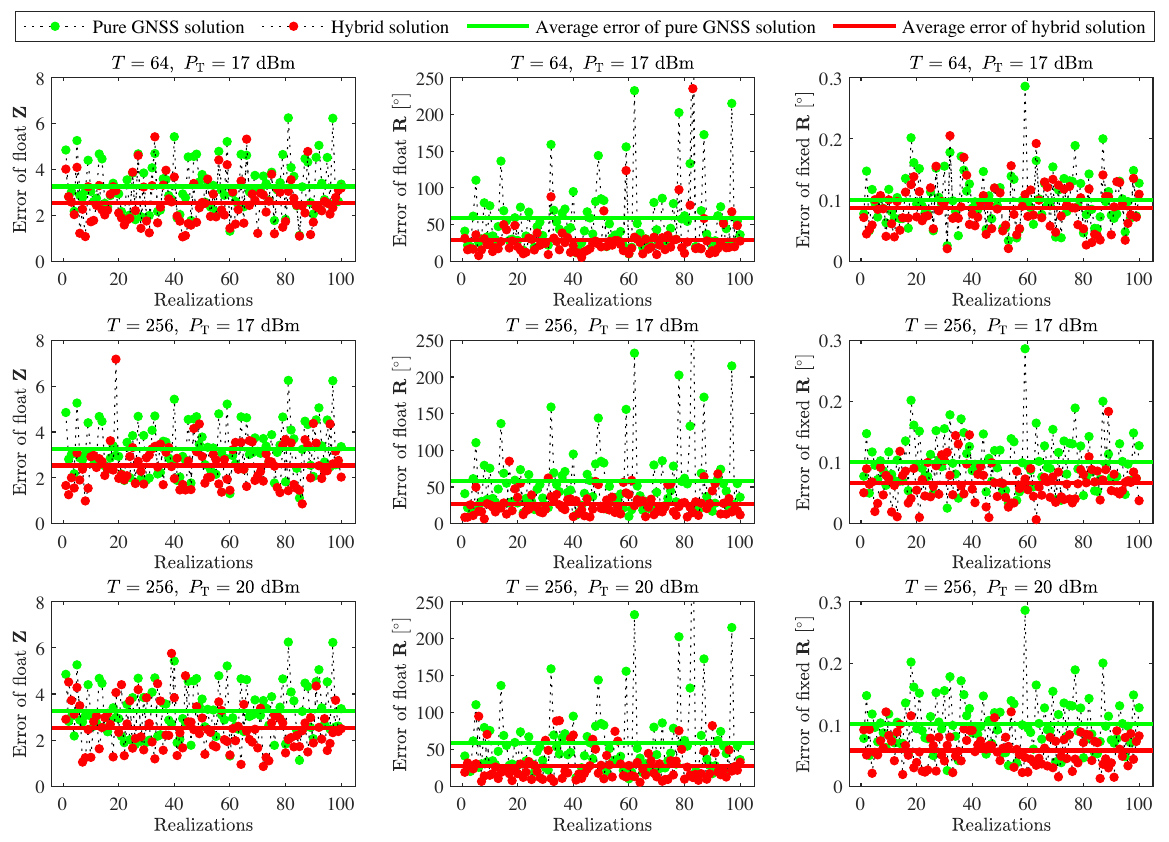}
    \caption{ 
		Estimation errors of the float ambiguity $\Zm$,
the float attitude $\Rm$, and the fixed attitude $\Rm$ for the pure \ac{gnss} solution and the hybrid solution (aided by a single \ac{5g} \ac{bs}) under setups: (i) $T=64,\ P_\mathrm{T}=\unit[17]{dBm}$; (ii) $T=256,\ P_\mathrm{T}=\unit[17]{dBm}$; (iii) $T=256,\ P_\mathrm{T}=\unit[20]{dBm}$.
      }
    \label{fig_2}
 \end{figure}

We first consider a scenario with a single \ac{5g} \ac{bs} to aid \ac{gnss} attitude determination. In this case, a standalone \ac{5g} attitude estimation is impossible.
Note that the performance of the \ac{5g}-based estimation usually depends on the transmission number $T$ and the signal power $P_\mathrm{T}$, whose impact will be tested in our simulations.
Fig.~\ref{fig_2} shows the estimation errors of the float ambiguity $\Zm$,
the float attitude $\Rm$, and the fixed attitude $\Rm$ for the pure \ac{gnss} solution and the hybrid solution, under 3 different setups of the \ac{5g} transmissions $T$ and signal power $P_\mathrm{T}$, i.e.: (i) $T=64,\ P_\mathrm{T}=\unit[17]{dBm}$; (ii) $T=256,\ P_\mathrm{T}=\unit[17]{dBm}$; (iii) $T=256,\ P_\mathrm{T}=\unit[20]{dBm}$.
The presented errors are calculated as the Frobenius norm of the difference between the estimated unknown matrices and the true ones.
We plot the average errors computed from 100 independent realizations for each setup, and the corresponding average errors are computed and plotted.
It can be clearly seen that even a single \ac{5g} \ac{bs} can help reduce the estimation errors of the \ac{gnss} attitude determination, in both the float solution and the fixed solution.
In addition, the estimation performance can be effectively improved by increasing the \ac{5g} transmissions $T$ and the \ac{5g} signal power $P_\mathrm{T}$.

\begin{figure}[t]
    \centering
    \includegraphics[width=4.5in]{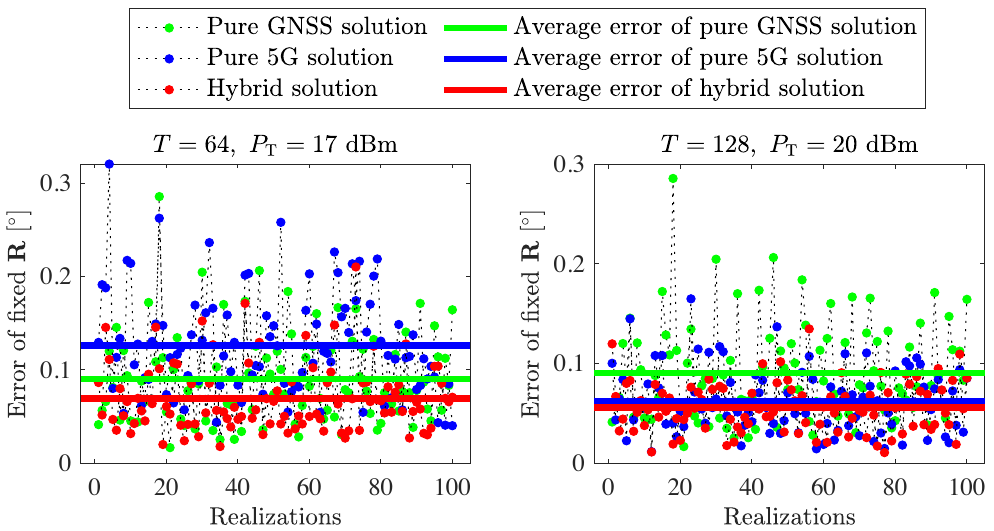}
    \caption{ 
		Estimation errors of the fixed attitude solution (of $\Rm$) for the pure \ac{gnss} solution, the pure \ac{5g} solution with 3 \acp{bs}, and the hybrid solution (aided by 3 \ac{5g} \acp{bs}) in 2 different setups: (i) $T=64,\ P_\mathrm{T}=\unit[17]{dBm}$; (ii) $T=128,\ P_\mathrm{T}=\unit[20]{dBm}$.
      }
    \label{fig_3}
    \vspace{0.4em}
    \centering
    \includegraphics[width=3.5in]{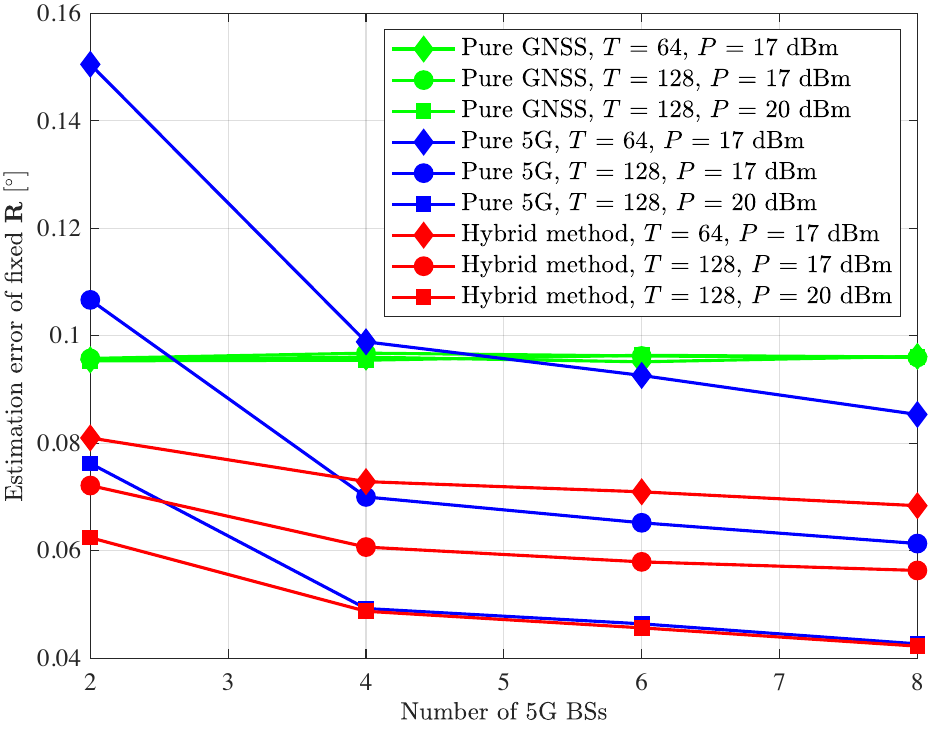}
    \caption{ 
		The average estimation errors of the pure \ac{gnss} solution, the pure \ac{5g} solution, and the hybrid solution versus the number of \ac{5g} \acp{bs}.
      }
    \label{fig_4}
 \end{figure}
 
 Next, we test the estimation performance when the number of \ac{5g} \acp{bs} is increased to 3. In these cases, a standalone \ac{5g} attitude determination is feasible.
 Therefore, we assess and compare the performance of the pure \ac{gnss} solution, the pure \ac{5g} solution, and the hybrid solution, to demonstrate the superiority of the proposed hybrid solution.
 Fig.~\ref{fig_3} demonstrates the estimation errors of the fixed attitude $\Rm$ for the pure \ac{gnss} solution, the pure \ac{5g} solution, and the hybrid solution under 2 different setups: (i) $T=64,\ P_\mathrm{T}=\unit[17]{dBm}$; (ii) $T=128,\ P_\mathrm{T}=\unit[20]{dBm}$.
For each setup, the results from 100 independent realizations are presented along with the average errors.
 We can observe that the hybrid solution outperforms the two standalone methods under various setups (when the pure \ac{gnss} solution outperforms the pure \ac{5g} solution or vice versa), which reveals that despite the feasibility of the pure \ac{gnss} and \ac{5g} solutions, the hybrid method can provide a further performance improvement.

 \begin{table}[t]
	\begin{minipage}[t]{0.45\textwidth}
	\renewcommand{\arraystretch}{1.3}
		\centering
		\caption{Success Rate of the Ambiguity Resolution ($\sigma~=~\unit[0.03]{m},\ T=64$)}
		\setlength{\tabcolsep}{2.5mm}
		\begin{tabular}{|c|c|c|c|c|c|}
		\hline
		\diagbox{$N+1$}{$L$} & 0 & 1 & 2 & 3 & 4 \\
		\hline
		\hline
		5 & 0\% & 0\% & 0\% & 14\% & 60\%\\
		\hline
		6 & 0\% & 0\% & 0\% & 26\% & 63\%\\
		\hline
		7 & 0\% & 0\% & 0\% & 27\% & 67\%\\
		\hline
		8 & 0\% & 0\% & 0\% & 30\% & 68\%\\
		\hline
		\end{tabular}
		\label{tab2}
	\end{minipage}
	\qquad
	\begin{minipage}[t]{0.45\textwidth}
	\renewcommand{\arraystretch}{1.3}
		\centering
		\caption{Estimation Error of the Fixed Attitude ($^\circ$) ($\sigma~=~\unit[0.03]{m},\ T=64$)}
		\setlength{\tabcolsep}{2.5mm}
		\begin{tabular}{|c|c|c|c|c|c|}
		\hline
		\diagbox{$N+1$}{$L$} & 0 & 1 & 2 & 3 & 4 \\
		\hline
		\hline
   		5 & 95.41 & 80.99  &  12.81  &  0.38  &  0.10 \\
   		\hline
  		6 & 92.26 & 81.37  &  12.85  &  0.31  &  0.09 \\
  		\hline
  		7 & 90.76 & 79.25  &  9.35  &  0.26  &  0.09  \\
  		\hline
  		8 & 87.70 & 79.13  &  9.31  &  0.19  &  0.09 \\
		\hline
		\end{tabular}
		\label{tab3}
	\end{minipage}
	\\ \vspace{1em} \\ 
	\begin{minipage}[t]{0.45\textwidth}
	\renewcommand{\arraystretch}{1.3}
		\centering
		\caption{Success Rate of the Ambiguity Resolution ($\sigma~=~\unit[0.03]{m},\ T=512$)}
		\setlength{\tabcolsep}{2.5mm}
		\begin{tabular}{|c|c|c|c|c|c|}
		\hline
		\diagbox{$N+1$}{$L$} & 0 & 1 & 2 & 3 & 4 \\
		\hline
		\hline
		5 & 0\% & 0\% & 0\% & 16\% & 70\%\\
		\hline
		6 & 0\% & 0\% & 0\% & 21\% & 71\%\\
		\hline
		7 & 0\% & 0\% & 0\% & 28\% & 75\%\\
		\hline
		8 & 0\% & 0\% & 0\% & 32\% & 75\%\\
		\hline
		\end{tabular}
		\label{tab4}
	\end{minipage}
	\qquad
	\begin{minipage}[t]{0.45\textwidth}
	\renewcommand{\arraystretch}{1.3}
		\centering
		\caption{Estimation Error of the Fixed Attitude ($^\circ$) ($\sigma~=~\unit[0.03]{m},\ T=512$)}
		\setlength{\tabcolsep}{2.5mm}
		\begin{tabular}{|c|c|c|c|c|c|}
		\hline
		\diagbox{$N+1$}{$L$} & 0 & 1 & 2 & 3 & 4 \\
		\hline
		\hline
   		5 & 93.09 & 80.69  & 12.69  &  0.08  &  0.04\\
   		\hline
   		6 & 86.54 & 80.31  &  6.84  &  0.06  &  0.03\\
   		\hline
   		7 & 87.90 & 80.28  &  7.52  &  0.06  &  0.03\\
   		\hline
   		8 & 85.02 & 81.12  &  2.44  &  0.06  &  0.03\\
		\hline
		\end{tabular}
		\label{tab5}
	\end{minipage}
	\\ \vspace{1em} \\ 
	\begin{minipage}[t]{0.45\textwidth}
	\renewcommand{\arraystretch}{1.3}
		\centering
		\caption{Success Rate of the Ambiguity Resolution ($\sigma~=~\unit[0.003]{m},\ T=512$)}
		\setlength{\tabcolsep}{1.5mm}
		\begin{tabular}{|c|c|c|c|c|c|}
		\hline
		\diagbox{$N+1$}{$L$} & 0 & 1 & 2 & 3 & 4 \\		
		\hline
		\hline
		5 & 14\% & 100\% & 100\% & 100\% & 100\%\\
		\hline
		6 & 98\% & 100\% & 100\% & 100\% & 100\%\\
		\hline
		7 & 100\% & 100\% & 100\% & 100\% & 100\%\\
		\hline
		8 & 100\% & 100\% & 100\% & 100\% & 100\%\\
		\hline
		\end{tabular}
		\label{tab6}
	\end{minipage}
	\qquad
	\begin{minipage}[t]{0.45\textwidth}
	\renewcommand{\arraystretch}{1.3}
		\centering
		\caption{Estimation Error of the Fixed Attitude ($^\circ$) ($\sigma~=~\unit[0.003]{m},\ T=512$)}
		\setlength{\tabcolsep}{2.75mm}
		\begin{tabular}{|c|c|c|c|c|c|}
		\hline
		\diagbox{$N+1$}{$L$} & 0 & 1 & 2 & 3 & 4 \\
		\hline
		\hline
    	5 & 30.00 & 0.11  &  0.06  &  0.06  &  0.03\\
    	\hline
    	6 & 1.33 & 0.11  &  0.06  &  0.05  &  0.03\\
    	\hline
    	7 & 0.21 & 0.10  &  0.06  &  0.05  &  0.02\\
    	\hline
    	8 & 0.22 & 0.09  &  0.06  &  0.04  &  0.02\\
		\hline
		\end{tabular}
		\label{tab7}
	\end{minipage}
\end{table}

 In Fig.~\ref{fig_4}, we further evaluate the root mean square errors (RMSEs) of the pure \ac{gnss} solution, the pure \ac{5g} solution, and the hybrid solution, for different numbers of \ac{5g} \acp{bs} $L=\{2,4,6,8\}$.
 The RMSE at each point is computed from 10000 independent trials.
Three setups are tested, i.e.: (i) $T=64,\ P_\mathrm{T}=\unit[17]{dBm}$; (ii) $T=128,\ P_\mathrm{T}=\unit[17]{dBm}$; (iii) $T=128,\ P_\mathrm{T}=\unit[20]{dBm}$.
 We can see that the performance of both the pure \ac{5g} solution and the hybrid solution improve as the number of \ac{5g} \acp{bs} increases, while the performance of the pure \ac{gnss} solution remains unchanged.
 In all the tested cases, the hybrid solution outperforms the other two methods.
 However, we observe that the gap between the errors of the pure \ac{5g} solution and the hybrid solution decreases with any increase in the number of \ac{5g} \acp{bs} $L$, the number of the \ac{5g} transmissions $T$, or the \ac{5g} signal power $P_\mathrm{T}$. This phenomenon indicates a performance saturation of the hybrid solution. Specifically, we can conclude that: (i) By increasing the number of \ac{5g} \acp{bs} $L$, the number of the \ac{5g} transmissions $T$, and/or the \ac{5g} transmission power $P_\mathrm{T}$, the performance of the pure \ac{5g} solution can be improved; (ii) the hybrid method cannot provide a significant performance improvement when the pure \ac{5g} solution is already much more accurate than the pure \ac{gnss} solution.
 However, in practical situations, maintaining \ac{5g} connection between the user platform and a large number of 5G \acp{bs} (with large power and transmission allocation) may not be achievable, where our proposed method can provide a significant performance improvement.

Finally, Table~\ref{tab2}--Table~\ref{tab7} demonstrate the success rate of the ambiguity resolution and the average estimation error of the fixed attitude of the proposed hybrid method for different numbers of satellites and different numbers of \ac{5g} \acp{bs}. 
Note that the case where $L=0$ coincides with the pure \ac{gnss} solution.
In these trials, 3 setups are tested: (i) $\sigma~=~\unit[0.03]{m},\ T=64$; (ii) $\sigma~=~\unit[0.03]{m},\ T=512$; (iii) $\sigma~=~\unit[0.003]{m},\ T=512$.
First, for high carrier-phase standard deviations $\sigma$ (e.g., $\sigma=0.03$), the standalone \ac{gnss} method cannot resolve the integer ambiguity at all with all the tested number of satellites.
However, when there are enough \ac{5g} \acp{bs} involved (e.g., $L=\{3,4\}$), the success rate becomes greater than zero, and the more \ac{5g} \acp{bs} we utilize, the higher the ambiguity resolution success rate. 
Consequently, it is naturally observed that the average estimation error is lowered with the increment of the ambiguity resolution success rate.
By comparing Table~\ref{tab3} and Table~\ref{tab5}, we observe that increasing the number of \ac{5g} transmissions also helps lower estimation errors.
Furthermore, the results in Table~\ref{tab6} and Table~\ref{tab7} show that even with successfully resolved integer ambiguities, increasing the number of \ac{5g} \acp{bs} and the number of satellites is helpful in enhancing the attitude estimation accuracy.

\section{Conclusion}
This paper formulated and solved the hybrid \ac{5g}-\ac{gnss} attitude determination problem, which is a promising solution in deep urban areas.
The observation model of the \ac{gnss} and \ac{5g} systems are described, where the hybrid observables consist of the GNSS pseudo-ranges, GNSS carrier phases, and \ac{5g} \acp{aoa}.
By rigorously incorporating these observations, a constrained weighted \ac{ls} problem is constructed.
This problem is solved by first obtaining an unconstrained float solution, then applying the MC-LAMBDA procedure to resolve integer ambiguity and obtain the fixed solution.
The proposed hybrid solution is compared with the pure \ac{gnss} and the pure \ac{5g} solutions, which showed the superiority of the proposed hybrid solution in both enhancing the ambiguity resolution success rate and improving the estimation accuracy.

\section*{ACKNOWLEDGEMENTS}

This publication is based upon work supported by the King Abdullah University of Science and Technology (KAUST) Office of Sponsored Research (OSR) under Award No. ORA-CRG2021-4695.

\bibliographystyle{apalike}
\bibliography{references}

\end{document}

%% file: macros.tex
\newcommand{\TT}{\mathsf{T}}

\newcommand{\bv}{{\bf b}}

\newcommand{\hv}{{\bf h}}

\newcommand{\nv}{{\bf n}}

\newcommand{\pv}{{\bf p}}

\newcommand{\tv}{{\bf t}}

\newcommand{\yv}{{\bf y}}
\newcommand{\zv}{{\bf z}}

\newcommand{\Am}{{\bf A}}
\newcommand{\Bm}{{\bf B}}

\newcommand{\Dm}{{\bf D}}
\newcommand{\Em}{{\bf E}}
\newcommand{\Fm}{{\bf F}}
\newcommand{\Gm}{{\bf G}}

\newcommand{\Mm}{{\bf M}}
\newcommand{\Nm}{{\bf N}}

\newcommand{\Pm}{{\bf P}}
\newcommand{\Qm}{{\bf Q}}
\newcommand{\Rm}{{\bf R}}

\newcommand{\Ym}{{\bf Y}}
\newcommand{\Zm}{{\bf Z}}

\newcommand{\Bt}{{\rm B}}

\newcommand{\Gt}{{\rm G}}

\newcommand{\Lt}{{\rm L}}

\newcommand{\Rt}{{\rm R}}

\newcommand{\Ut}{{\rm U}}

\newcommand{\Sigmam}{\hbox{\boldmath$\Sigma$}}

\newcommand{\Xim}{\hbox{\boldmath$\Xi$}}

%% file: IONconf_main.bbl
\begin{thebibliography}{}

\bibitem[3GPP, 2023]{tr2370086}
3GPP (2023).
\newblock {{3GPP TR 23.700-86 V18.0.0:} Study on Architecture Enhancement to
  support ranging based services and sidelink positioning (Release 18)}.

\bibitem[Abu-Shaban et~al., 2018]{Abu2018Error}
Abu-Shaban, Z., Zhou, X., Abhayapala, T., Seco-Granados, G., and Wymeersch, H.
  (2018).
\newblock Error bounds for uplink and downlink {3D} localization in {5G}
  millimeter wave systems.
\newblock {\em IEEE Transactions on Wireless Communications}, 17(8):4939--4954.

\bibitem[Alletto et~al., 2016]{Alletto2016Indoor}
Alletto, S., Cucchiara, R., Del~Fiore, G., Mainetti, L., Mighali, V., Patrono,
  L., and Serra, G. (2016).
\newblock An indoor location-aware system for an {IoT}-based smart museum.
\newblock {\em IEEE Internet of Things Journal}, 3(2):244--253.

\bibitem[Angrisano et~al., 2010]{angrisano2010gnss}
Angrisano, A. et~al. (2010).
\newblock {GNSS/INS} integration methods.
\newblock {\em Dottorato di ricerca (PhD) in Scienze Geodetiche e Topografiche
  Thesis, Universita’degli Studi di Napoli PARTHENOPE, Naple}, 21.

\bibitem[Chen et~al., 2022]{Chen2022Tutorial}
Chen, H., Sarieddeen, H., Ballal, T., Wymeersch, H., Alouini, M.-S., and
  Al-Naffouri, T.~Y. (2022).
\newblock A tutorial on terahertz-band localization for {6G} communication
  systems.
\newblock {\em IEEE Communications Surveys \& Tutorials}, 24(3):1780--1815.

\bibitem[Chen et~al., 2023]{chen2023multi}
Chen, H., Zheng, P., Keskin, M.~F., Al-Naffouri, T.~Y., and Wymeersch, H.
  (2023).
\newblock {Multi-RIS-enabled 3D} sidelink positioning.
\newblock {\em preprint arXiv:2302.12459}.

\bibitem[del Peral-Rosado et~al., 2018]{del2018Survey}
del Peral-Rosado, J.~A., Raulefs, R., López-Salcedo, J.~A., and Seco-Granados,
  G. (2018).
\newblock Survey of cellular mobile radio localization methods: From {1G to
  5G}.
\newblock {\em IEEE Communications Surveys \& Tutorials}, 20(2):1124--1148.

\bibitem[Di~Taranto et~al., 2014]{Di2014Location}
Di~Taranto, R., Muppirisetty, S., Raulefs, R., Slock, D., Svensson, T., and
  Wymeersch, H. (2014).
\newblock Location-aware communications for {5G} networks: How location
  information can improve scalability, latency, and robustness of {5G}.
\newblock {\em IEEE Signal Processing Magazine}, 31(6):102--112.

\bibitem[Douik et~al., 2020]{Douik2020Precise}
Douik, A., Liu, X., Ballal, T., Al-Naffouri, T.~Y., and Hassibi, B. (2020).
\newblock Precise {3-D GNSS} attitude determination based on {Riemannian}
  manifold optimization algorithms.
\newblock {\em IEEE Transactions on Signal Processing}, 68:284--299.

\bibitem[Falco et~al., 2017]{falco2017loose}
Falco, G., Pini, M., and Marucco, G. (2017).
\newblock Loose and tight {GNSS/INS} integrations: Comparison of performance
  assessed in real urban scenarios.
\newblock {\em Sensors}, 17(2):255.

\bibitem[Giorgi and Teunissen, 2010]{giorgi2010carrier}
Giorgi, G. and Teunissen, P.~J. (2010).
\newblock Carrier phase {GNSS} attitude determination with the multivariate
  constrained {LAMBDA} method.
\newblock In {\em IEEE Aerospace Conference}.

\bibitem[Groves, 2013]{groves2013gnss}
Groves, P. (2013).
\newblock {GNSS} solutions: Multipath vs. {NLOS} signals. how does
  non-line-of-sight reception differ from multipath interference.
\newblock {\em Inside GNSS Magazine}, 8(6):40--42.

\bibitem[He et~al., 2021]{He2021Channel}
He, J., Wymeersch, H., and Juntti, M. (2021).
\newblock Channel estimation for {RIS}-aided mmwave {MIMO} systems via atomic
  norm minimization.
\newblock {\em IEEE Transactions on Wireless Communications}, 20(9):5786--5797.

\bibitem[Horn and Johnson, 2012]{horn2012matrix}
Horn, R.~A. and Johnson, C.~R. (2012).
\newblock {\em Matrix analysis}.
\newblock Cambridge university press.

\bibitem[Jiang et~al., 2021]{Jiang2021OmniTrack}
Jiang, C., He, Y., Zheng, X., and Liu, Y. (2021).
\newblock Omnitrack: Orientation-aware {RFID} tracking with centimeter-level
  accuracy.
\newblock {\em IEEE Transactions on Mobile Computing}, 20(2):634--646.

\bibitem[Kay, 1993]{kay1993fundamentals}
Kay, S.~M. (1993).
\newblock {\em Fundamentals of statistical signal processing: estimation
  theory}.
\newblock Prentice-Hall, Inc.

\bibitem[Liu, 2023]{liu2023gnss}
Liu, X. (2023).
\newblock {\em {GNSS} Localization and Attitude Determination via Optimization
  Techniques on {Riemannian} Manifolds}.
\newblock PhD thesis.

\bibitem[Liu et~al., 2022a]{liu2022gnss}
Liu, X., Ballal, T., Ahmed, M., and Al-Naffouri, T.~Y. (2022a).
\newblock A {GNSS} attitude determination algorithm using optimization
  techniques on {Riemannian} manifolds.
\newblock In {\em Proceedings of the 35th International Technical Meeting of
  the Satellite Division of The Institute of Navigation (ION GNSS+ 2022)},
  pages 2064--2073.

\bibitem[Liu et~al., 2023]{Liu2022Instantaneous}
Liu, X., Ballal, T., Ahmed, M., and Al-Naffouri, T.~Y. (2023).
\newblock Instantaneous {GNSS} ambiguity resolution and attitude determination
  via {Riemannian} manifold optimization.
\newblock {\em IEEE Transactions on Aerospace and Electronic Systems},
  59(3):3296--3312.

\bibitem[Liu et~al., 2019]{liu2019gnss}
Liu, X., Ballal, T., and Al-Naffouri, T.~Y. (2019).
\newblock {GNSS}-based localization for autonomous vehicles: Prospects and
  challenges.
\newblock In {\em 27th European Signal Processing Conference (EUSIPCO)}.

\bibitem[Liu et~al., 2020]{9110131}
Liu, X., Ballal, T., and Al-Naffouri, T.~Y. (2020).
\newblock {GNSS} attitude determination using a constrained wrapped least
  squares approach.
\newblock In {\em 2020 IEEE/ION Position, Location and Navigation Symposium
  (PLANS)}, pages 1135--1139.

\bibitem[Liu et~al., 2022b]{9837939}
Liu, X., Ballal, T., Chen, H., and Al-Naffouri, T.~Y. (2022b).
\newblock Constrained wrapped least squares: A tool for high-accuracy {GNSS}
  attitude determination.
\newblock {\em IEEE Transactions on Instrumentation and Measurement}, 71:1--15.

\bibitem[Shahmansoori et~al., 2018]{Shahmansoori2018Position}
Shahmansoori, A., Garcia, G.~E., Destino, G., Seco-Granados, G., and Wymeersch,
  H. (2018).
\newblock Position and orientation estimation through millimeter-wave {MIMO} in
  {5G} systems.
\newblock {\em IEEE Transactions on Wireless Communications}, 17(3):1822--1835.

\bibitem[Teunissen, 2012]{teunissen2012affine}
Teunissen, P. (2012).
\newblock The affine constrained {GNSS} attitude model and its multivariate
  integer least-squares solution.
\newblock {\em Journal of geodesy}, 86(7):547--563.

\bibitem[Teunissen and Tiberius, 1994]{teunissen1994integer}
Teunissen, P. and Tiberius, C. (1994).
\newblock Integer least-squares estimation of the {GPS} phase ambiguities.
\newblock In {\em Proceedings of the International Symp. On Kinematic Systems
  in Geodesy, Aug, 30-Sept. 2, 1994, Banff, Canada, 11 pp}.

\bibitem[Teunissen and Kleusberg, 2012]{teunissen2012gps}
Teunissen, P.~J. and Kleusberg, A. (2012).
\newblock {\em GPS for Geodesy}.
\newblock Springer Science \& Business Media.

\bibitem[Teunissen and Montenbruck, 2017]{teunissen2017springer}
Teunissen, P.~J. and Montenbruck, O. (2017).
\newblock {\em Springer handbook of global navigation satellite systems},
  volume~10.
\newblock Springer.

\bibitem[Wen et~al., 2021]{wen2021factor}
Wen, W., Pfeifer, T., Bai, X., and Hsu, L.-T. (2021).
\newblock Factor graph optimization for {GNSS/INS} integration: A comparison
  with the extended {Kalman} filter.
\newblock {\em NAVIGATION: Journal of the Institute of Navigation},
  68(2):315--331.

\bibitem[Zheng et~al., 2022]{Zheng2022Coverage}
Zheng, P., Ballal, T., Chen, H., Wymeersch, H., and Al-Naffouri, T.~Y. (2022).
\newblock Coverage analysis of joint localization and communication in {THz}
  systems with {3D} arrays.
\newblock {\em TechRxiv preprint}.

\bibitem[Zheng et~al., 2023a]{Zheng2023JrCUP}
Zheng, P., Chen, H., Ballal, T., Valkama, M., Wymeersch, H., and Al-Naffouri,
  T.~Y. (2023a).
\newblock {JrCUP}: Joint {RIS} calibration and user positioning for {6G}
  wireless systems.
\newblock {\em preprint arXiv:2304.00631}.

\bibitem[Zheng et~al., 2023b]{Zheng2023Misspecified}
Zheng, P., Chen, H., Ballal, T., Wymeersch, H., and Al-Naffouri, T.~Y. (2023b).
\newblock Misspecified {Cramér-Rao} bound of {RIS}-aided localization under
  geometry mismatch.
\newblock In {\em 2023 IEEE International Conference on Acoustics, Speech and
  Signal Processing (ICASSP)}.

\bibitem[Zheng et~al., 2023c]{Zheng20235G}
Zheng, P., Liu, X., Ballal, T., and Al-Naffouri, T.~Y. (2023c).
\newblock {5G}-aided {RTK} positioning in {GNSS}-deprived environments.
\newblock In {\em 31th European Signal Processing Conference (EUSIPCO)}.

\end{thebibliography}
